# Isothermal Annealing Effects on β-Relaxations and Crystallization Behaviors in Amorphous GeTe

Arune Makareviciute[1], Qun Yang[2,3], Tomoki Fujita[1], Oliver Gross[4], Nico Neuber[4], Maximilian Frey[4], Jens Moesgaard[1], Cecile Chaxel[1], Julian Pries[5], Mads Ry Vogel Jørgensen[1,6], Frederik Holm Gjørup[1,6], Matthias Wuttig[5], Hai-bin Yu[2*], Jiangjing Wang[5,7*], Shuai Wei[*1,8]

[1]*Department of Chemistry, Aarhus University, 8000 Aarhus C, Denmark.*

[2]*Wuhan National High Magnetic Field Center and School of Physics, Huazhong University of Science and Technology, Wuhan 430074, Hubei, China.*

[3]*College of Physics and Electronic Engineering, Chongqing Normal University, Chongqing 401331, China.*

[4]*Chair of Metallic Materials, Saarland University, Campus C6.3, 66123 Saarbrücken, Germany*

[5]*Institute of Physics (IA), RWTH Aachen University, Aachen, 52056, Germany.*

[6]*MAX IV Laboratory, Lund University, 224 84 Lund, Sweden.*

[7]*Center for Alloy Innovation and Design (CAID), State Key Laboratory for Mechanical Behavior of Materials, Xi'an Jiaotong University, Xi'an, 710049, China.*

[8]*CENSEMAT Centre for Energy Materials Research, 8000 Aarhus, Denmark*

**Abstract**

A secondary (β) relaxation process is often the dominant source of atomic dynamics below $T_g$ in many glass forming systems. Recent studies reported the presence of β-relaxations in amorphous phase-change materials (PCMs) and showed that suppressing the β-relaxation via annealing in $Ge_{15}Sb_{85}$ can effectively slow down its crystallization kinetics. Yet, when Sb is replaced by Te, similar annealing protocol has little effect on the Te-rich alloy $Ge_{15}Te_{85}$. Here, we investigate amorphous GeTe that is a Sb-free PCM, but with faster crystallization kinetics than $Ge_{15}Te_{85}$. Using powder mechanical dynamic spectroscopy, we observe a clear reduction of the excess-wing in the loss modulus upon isothermal annealing, indicating a suppression of its β-relaxation. Ultrafast calorimetric analysis and time-resolved optical reflectivity measurements show that, whereas as-deposited GeTe exhibit stochastic crystallization behaviors, annealed samples crystallize more slowly with reduced stochasticity. Synchrotron X-ray scattering experiments reveal reinforced Peierls-like distortions in the amorphous structure after annealing, and demonstrate that, even if annealing introduces nucleation sites, it nonetheless slows down crystallization kinetics. These finding suggests that, in annealed GeTe, crystallization is limited by crystal growth rate, which is retarded through the suppression of β-relaxation.

*haibinyu@hust.edu.cn (HBY), j.wang@xjtu.edu.cn (JJW), shuai.wei@chem.au.dk (SW)

## 1. Introduction

Relaxation processes are intrinsic for all glasses and liquids. In studies of metallic, molecular, and polymer glasses, a secondary (β) relaxation has been often observed in their glassy states[1–3]. As a main source of glass dynamics, where the primary (α-) relaxation is frozen-in below $T_g$, the β-relaxation is associated with local fast atomic motion embedded in slow mobility matrix[1,4], which is often argued to result from the heterogenous structures of glasses[5–7]. The local fast atomic motion is considered as reversible cooperative atomic rearrangement but may transform to irreversible motion as the temperature approaches the glass transition, where α-relaxation is activated[8,9]. β-relaxations are shown important for their amorphous stabilities, crystallization, and mechanical properties[1,7].

Amorphous PCMs (e.g. GeTe, $Ge_2Sb_2Te_5$, and $Ge_{15}Sb_{85}$) belong to extremely poor glass formers with a critical cooling rate of the order of $10^{10}$ K/s for vitrification[10,11]. The capability of being reversible and fast switching between amorphous and crystalline phases, as well as the resulting strong property contrast between the two phases, make them promising material candidates for nonvolatile data storage, neuromorphic, and photonic computing applications[12–14]. The poor glass forming ability makes it difficult to synthesize bulk samples required for conventional dynamic mechanical spectroscopy (DMS) measurements[15,16]. Recent studies, using powder DMS (PMS), overcame the sample geometry problem and probed the viscoelastic loss modulus ($E''$) in typical PCMs and non-PCM chalcogenide glasses[15]; only in PCMs, excess wings in $E''$ have been observed, indicating the presence of β-relaxations[15]. An annealing experiment at $0.9T_g$ in amorphous $Ge_{15}Sb_{85}$ was shown to suppress the β-relaxation, which slowed down crystallization kinetics[7]. $Ge_{15}Sb_{85}$ is a near-eutectic composition[17]. Replacing Sb by Te leads to the eutectic composition $Ge_{15}Te_{85}$ which exhibits vanishingly small β-relaxation and nearly no annealing effect[7]. However, the equal-atomic GeTe shows a pronounced β-relaxation[15], resembling that of $Ge_{15}Sb_{85}$. The crystallization of $Ge_{15}Sb_{85}$ is known as growth dominated[18]. A growth-dominated crystallization is also reported in GeTe[19]. A previous ab initio computational work of aging in amorphous GeTe suggested that the structural evolution during aging is characterized by reinforced Peierls-like distortions in local octahedral motifs and reduced volume fraction of tetrahedral motifs[20], although experimental structural evidence has been missing. The questions arise: how annealing affects the β-relaxation and crystallization in GeTe and their associated structural changes, as well as how their behavior differs from or resembles that of $Ge_{15}Sb_{85}$.

In this work, we combine complementary techniques to investigate the effects of thermal annealing on β-relaxation and crystallization in amorphous GeTe. PMS was employed to characterize the β-relaxation in as-deposited and annealed samples below $T_g$. Ultrafast calorimetry and optically induced crystallization experiments were employed to probe crystallization kinetics after suppression of the β-relaxations by annealing. The structural evolution upon annealing was further examined using synchrotron X-ray scattering, enabling monitoring of the amorphous phase and partially crystallized phases. Together, these results provide a better understanding of how annealing modifies β-relaxation and crystallization in GeTe.

2. Experimental Methods

**Sample preparations:** The amorphous samples of GeTe were prepared by magnetron sputtering deposition with stoichiometric targets at a background pressure of $3\times10^{-3}$ mbar and argon flow 20 sccm. Thick layers of several micrometers were sputtered to obtain sufficient powder or flake samples for PMS, (F)DSC, and XRD measurements. Sample compositions measured by EDS were within ~ 1 at. % from the nominal composition (within EDS error range).

**Powder mechanical dynamic spectroscopy (PMS):** Prior to PMS testing, the exfoliated flakes were carefully milled into powders with a uniform particle size. For PMS measurements, a TA Q-800 dynamic mechanical analyzer (DMA) was employed in combination with a custom-designed powder container. Temperature-dependent relaxation spectra of powder samples were recorded in multi-frequency strain mode at discrete fixed testing frequencies (0.5, 1, 2, and 8 Hz), with a strain amplitude of 10 μm and a heating rate of 3 K/min, all conducted under an argon atmosphere. As previously described in Ref.[7], the absolute values obtained from PMS measurements do not directly reflect the intrinsic viscoelastic properties of the material, as they are strongly influenced by the physical state of the powder sample, including its mass and compactness. To ensure data reliability, each measurement used a consistent powder mass of 300 mg, and the torsional force applied to the powder clamp was maintained at 5 pounds. Furthermore, to enable quantitative comparison of relaxation behavior, all temperature-dependent $E''$ spectra were normalized. Annealing experiments were also performed on powder samples within the DMA chamber under argon protection. In this experiment, the chamber temperature was rapidly equilibrated to the target annealing temperature, followed by in-situ annealing at that temperature for 3 h, and then rapid cooling to room temperature. The relaxation spectra of annealed samples were obtained under identical testing conditions and parameters to those described above.

**Calorimetric measurements**: Differential scanning calorimetry (DSC) was employed to investigate crystallization kinetics in GeTe. A TA DSC25 device (from TA Instruments) was used, where a 3–5 mg powder sample produced by the sputtering method described above was encapsulated in a Tzero-type aluminum pan with a lid and placed in a furnace filled with argon. Eight different heating rates ranging from 3 to 150 K/min were applied to individual samples of both as-deposited and annealed GeTe powders. The temperature program included ramping from 40 to 400 °C and a rescan at the same temperatures. To expand the heating-rate range, Flash Differential Scanning Calorimetry (FDSC) was employed. Thermal

Analysis System Flash DSC 2+ (from Mettler Toledo) with MultiSTAR UFS 1 sensor chips was used. FDSC measurements were conducted using 22 different heating rates spanning from 20 to 12000 K/s for both as-deposited and annealed GeTe samples. For FDSC measurement, a thin flake of GeTe powder, weighing a few hundred nanograms, was positioned at the sensor center and heated under a nitrogen atmosphere. The temperature program for FDSC measurements included ramping from −10 to 450 °C and a rescan. Five measurements were conducted at each heating rate to ensure statistical reliability. Each scan underwent subtraction by its subsequent rescan of the crystallized sample. Thereafter, the crystallization exotherm was normalized to heating rate and sample mass. The crystallization peak temperature obtained from DSC and the averaged peak temperature from FDSC results were utilized to construct a Kissinger plot. Before data collection, both DSC and FDSC were temperature-calibrated at multiple heating rates using the indium melting peak. For DSC, the instrument software automatically applied temperature corrections derived from these calibrations. For FDSC, temperature corrections were calculated manually from the indium melting calibration and applied to each measurement.

**Optical laser induced crystallization**: The GeTe thin films of ~50 nm thickness were sputtered on silicon substrate and sandwiched between two layers of ZnS-SiO$_2$ (prepared by RF power). The thermal annealing was performed at 166°C in a glass tube with a heating rate of 5 K/min and holding time of 3 hrs under Ar atmosphere. The crystallization of the thin films was performed by applying focused laser pulses (pump laser, wavelength = 658 nm) with the power of ~40 mW and variable duration towards a series submicron region. The optical reflectivity measurements during phase transformation were performed with the low-intensity continuous-wave laser (wavelength = 639 nm) at the same position with the pump laser. The relative change in reflectance is $\Delta R=(R_i-R_f)/R_i$ where $R_i$ and $R_f$ are initial and final reflectivity, respectively. The fitting for minimum crystallization time $t_{min}$ is based on the Gompertz function, which is a sigmoid function that depicts the growth mode being slowest at the start and saturated for a relatively long time. The asymptote parameter $a$ is confined to 100 as the success rate should be no larger than 100%. The $\Delta R$ of 0.1 was used as the threshold for crystallization to begin. The standard deviation of probability of crystallization, $S_y$, can be evaluated by the equation: $S_y = \sqrt{\frac{1}{n-1}\sum_{i=1}^{n}(y_i - \bar{y})^2}$.

**Synchrotron X-ray scattering at DanMAX, MAX IV**: We performed *in-situ* synchrotron X-ray total scattering experiment at the DanMAX beamline of MAX IV. The photon energy of the monochromatic X-ray beam was selected as 35.00 keV, focused to 0.25 x 0.21 mm$^2$. Powder samples of amorphous GeTe were sealed in quartz capillaries with an inner diameter of 0.08 mm and a wall thickness of 0.01 mm and were aligned with the X-ray beam in Debye-Scherrer geometry. Two-dimensional scattering images were collected using a PILATUS3 X 2M CdTe detector, covering a $Q$-range from 0.5 Å$^{-1}$ to 16 Å$^{-1}$ with an exposure time of 1 s. The 2D diffraction images were azimuthally integrated using the MatFRAIA algorithm[21]. The sample temperature was controlled using Oxford Cryosystems Cryostream 800 plus. The first XRD dataset was collected during continuous heating of an as-deposited sample from room temperature (RT) to a targeted isothermal annealing temperature below the glass transition at a heating rate of 3 K min$^{-1}$, with a data acquisition rate of 1 Hz. Additional measurements were performed under isothermal conditions: two separate samples were annealed at 132 °C and 166 °C, respectively, each for 180 min, using the same acquisition rate. After isothermal annealing, the samples were cooled back to RT, where 10 diffraction patterns were recorded for each.

**Synchrotron X-ray scattering and pair distribution function analysis at DESY PETRAIII P02.1**: In parallel, we performed the total X-ray scattering measurements of an as-deposited and a pre-annealed sample (105 °C for 1 hr treated in a DSC) at DESY PETRAIII P02.1 for pair distribution function analysis. The photon energy and the beam size of the incident X-ray were selected as 59.83 keV and 0.5 × 0.5 mm$^2$. The powder/flake samples loaded in a quartz capillary (1 mm inner diameter and 0.01 mm wall thickness) were heated at 0.333 K/s in a custom-built ceramic heater under a constant argon flow. Measurements were carried out in a transmission mode with an exposure time of 1 s per frame, summed 10 frames for each pattern, using a Perkin Elmer XRD1621 CsI bonded amorphous silicon detector (2048 × 2048 pixels). Dark-subtracted two-dimensional X-ray diffraction patterns were integrated to obtain one-dimensional intensity profiles using the Fit2D[22]. The background was measured at room temperature and subtracted from the integrated intensity data. The data were processed using the PDFgetX2[23] to calculate total structure factors $S(Q)$ after background subtraction and corrections for absorption, fluorescence, and Compton scattering. The reduced pair distribution functions $G(r)$ were calculated by the Fourier transform of $S(Q)$ ($Q_{max}$ =16 Å$^{-1}$) in PDFgetX2 (see details in Ref.[24]).

3. Results
a. β-relaxations and thermal annealing effects

Powder mechanical spectroscopy (PMS) measurements allow for measuring viscoelastic loss modulus ($E''$) and storage modulus $E'$ of powder samples that lack bulk glass forming abilities required for conventional DMS measurements. $E''$ and $E'$ can be used to characterize the α- and β- relaxations in glasses at a given frequency $f$. Figure 1A shows the $E''$ measured at $f$ = 1 Hz for an as-deposited GeTe sample and a sample annealed at $T_{ann}$ = 166 °C (corresponding to ~ 0.93$T_g$ (K), where $T_g = T_\alpha$ as described below) for 3 hrs. For the as-deposited sample, a pronounced excess wing with a bump is observed at around 400 K below the main $α$-relaxation peak at $T_\alpha \approx$ 197 °C = 470 K. The excess wing is indicative of $β$-relaxation, while the $α$-relaxation peak corresponds to the glass transition, e.g. $T_g \approx T_\alpha$, for $f$ = 1 Hz. By contrast, the excess wing of the annealed sample is substantially reduced, indicating a suppressed β-relaxation. In addition, the α-relaxation peak of the annealed sample is slightly shifted to higher temperature ~209°C.

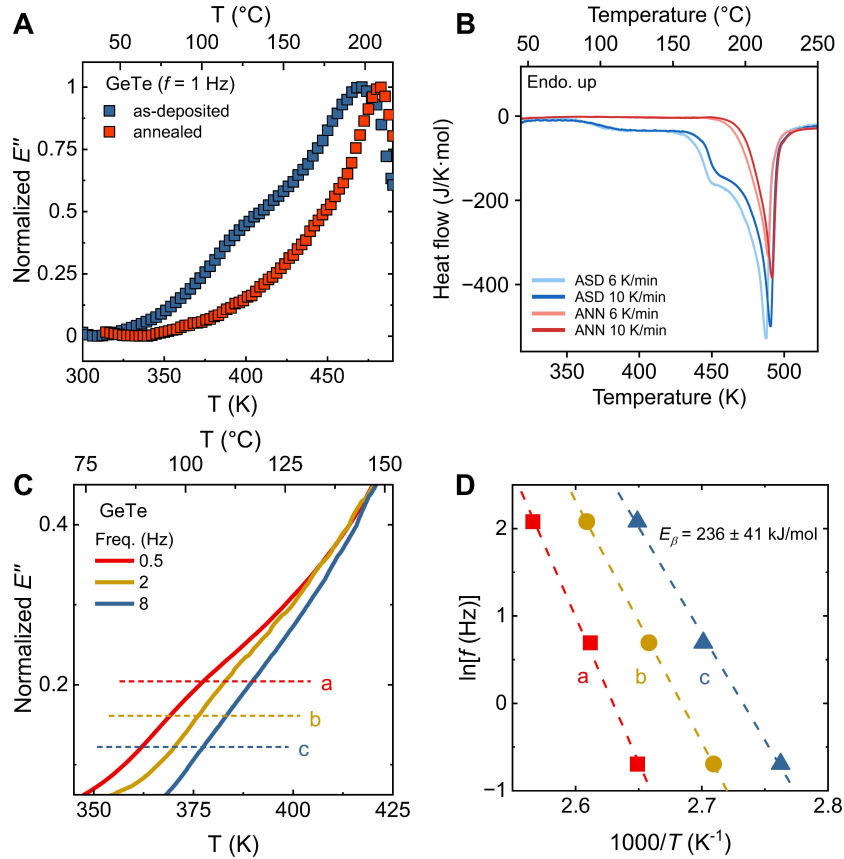

**Figure 1**: Effect of thermal annealing on *β*-relaxations in amorphous GeTe. (**A**) Normalized loss modulus $E''$ for as-deposited GeTe and annealed GeTe at 166 °C for 3 hrs. (**B**) DSC scans for deposited (ASD) and annealed (ANN) GeTe. (**C**) The normalized $E''$ of the as-deposited sample at different frequencies. (**D**) The fits to the frequency dependency of the $E''$ curve shift yields the activation energy of β-relaxation. The labels a, b, and c indicate the temperature points corresponding to the dashed lines in C.

Figure 1B shows the differential scanning calorimetry (DSC) scans of the as-deposited and annealed samples at 6 and 10 K/min. DSC scans allow for measuring heat flow (proportional to heat capacity). The as-deposited sample exhibits a pronounced exothermic event with a rather flat heat flow spanning from ~ 90°C to ~ 160°C. Upon further heating, the heat flow shows a stepwise drop above ~ 160 °C, which is followed by a large exothermic peak of crystallization (~215 °C). The exothermic event below 160 °C, corresponding to an enthalpy release of ~2.45 kJ/mol, reveals an enthalpy relaxation. Similar enthalpy relaxations have been observed in as-deposited amorphous $Ge_2Sb_2Te_5$ and $Ge_{15}Sb_{85}$[18,25], as well as also observed in those hypequenched glasses, where a large amount of enthalpy is frozen-in during rapid melt-quench vitrification[26–28]. However, the stepwise drop and plateau at ~ 160-190 °C can be either interpreted as enthalpy relaxation, or an initial step of crystallization. This point will be clarified by in-situ X-ray diffraction measurements in Section **c**. By contrast, the annealed sample shows no exothermic event before the crystallization peak sets in at ~ 200 °C. Note that the temperature range of enthalpy relaxation (~ 90°C to ~ 160°C) in the as-deposited is approximately the same range of the pronounced excess wing in $E''$ observed in PMS measurements. This implies that the enthalpy release and the β-relaxation are correlated processes within the present system GeTe.

Figure 1C-D shows the frequency dependent $E''$. Note that due to the limited quantities of amorphous samples that can be produced via magnetron sputtering deposition, the measurements were limited to a few frequencies, each requiring approximately 300 mg of material, which a substantial amount for this fabrication method. Since no clear peak temperature can be defined for the β-relaxation, we examine the $f$-dependent shift of the $E''$ curves for a few selected $E''$ values (Fig.1C). By fitting the *ln f* vs. *1000/T* (Fig.1D), we can estimate the averaged activation energy of the β-relaxation, $E_β$ = 236 ± 41 kJ/mol. Since $E_β$ has never been reported for other PCMs so far, we can only compare it with metallic glasses[29], where $E_β$ is reported ranging from 75 - 200 kJ/mol. $E_β$ of GeTe is clearly larger than that of metallic glasses. Moreover, $E_β/RT_g$ ≈ 60 for GeTe (assuming $T_g = T_α$ = 197 °C) is about 2 times larger than empirical coefficient ~26 ± 2 for metallic glasses[29].

b. **Crystallization kinetics**

A series of conventional DSC and ultrafast Flash DSC (FDSC) measurements were performed to examine the effect of annealing on crystallization kinetics. Figure 2A-B shows the heat flow of the as-deposited for the constant heating rate from 20 K/s to 12,000 K/s measured using FDSC. Figure 2C-D shows the heat flow of the annealed sample (166 °C, 3 hrs) for the same range of heating rates. DSC scans for lower heating rates are shown in Fig.S1. A combination of both DSC and FDSC enables us to cover a broad range of heating rates over six orders magnitude. The onset and peak temperatures of crystallization are plotted in Fig. 2E. Over the entire heating rate range, both onset and peak of the annealed sample exhibit a consistent shift toward higher temperatures relative to the as-deposited samples. This shift becomes increasingly pronounced at higher heating rates and suggests slower crystallization kinetics after annealing.

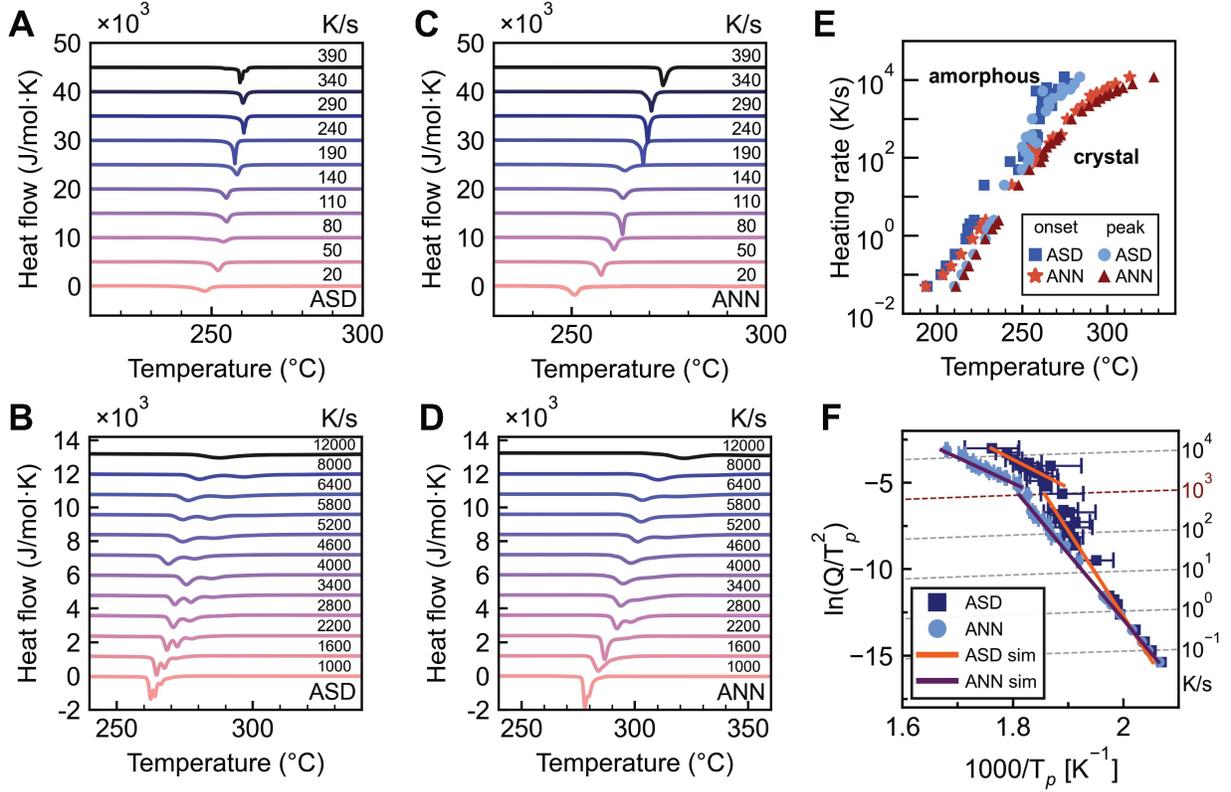

**Figure 2**: Calorimetric measurements for crystallization kinetics. The heat flow measured by FDSC scans for the as-deposited sample from 20 K/s to 390 K/s (**A**) and from 1000 to 12000 K/s (**B**). The heat flow for annealed samples (166 °C, 3 hrs) is shown in (**C**) and (**D**). (**E**) Kinetic map illustrating crystallization onset and peak variation based on heating rate for as-deposited and annealed samples. (**F**) The Kissinger plot for as-deposited (ASD) and annealed (ANN) GeTe. The plot combines results from conventional DSC and fast differential scanning calorimetry (FDSC) measurements over a heating rate range of 0.05 to 12,000 K/s. Error bars are shown for FDSC data points, which represent the average of five measurements. DSC points were obtained from single scans at each heating rate. Solid lines represent simulated Kissinger data, calculated separately for low ($Q \leq 10^3$ K/s) and high ($Q > 10^3$ K/s) heating rate regimes, where a change in slope is observed.

Furthermore, the crystallization peak temperatures ($T_p$) are analyzed and plotted in the Kissinger plot, $\ln(Q/T_p^2)$ versus $1/T_p$ for both as-deposited and annealed samples, as shown in Fig.2F. The data can be segmented into two regimes: low (up to $10^3$ K/s) and high (above $10^3$ K/s), based on an observed change in slope. Linear fits were used to extract parameters, an activation energy $E_K$ of 4.31 eV (low rates) and 1.49 eV (high rates) for as-deposited; 3.34 eV (low rates) and 1.35 eV (high rates) for annealed. With the fitted $E_K$, we performed numerical simulations based on the Avrami equation following the protocol documented in ref.[25] (see Supplementary text for details). The simulated Kissinger plots were generated for both regimes and are displayed as solid lines in the figure. From the simulations, we can calculate the crystalline volume fractions as a function of temperature at given heating rates for the annealed and as-deposited samples, shown in Fig.S2.

## c. Optically Induced Crystallization

Optical reflectivity measurements were used to independently probe crystallization and validate the annealing effect. Optical reflectivity contrast between the amorphous and crystalline phases enables direct tracking of the crystallization process, as changes in reflectivity correlate with the crystalline fraction formed within the amorphous GeTe matrix[12]. The thin films were exposed to laser pulses with the constant power of ~40 mW, and the duration range from $10^2$ ns to $10^5$ ns at different spots. We performed the laser measurements with 30 different pulse lengths $t$, and for each pulse length, 30 pump pulses were applied at different areas of the films (see details in *Methods* and also in Ref.[12]).

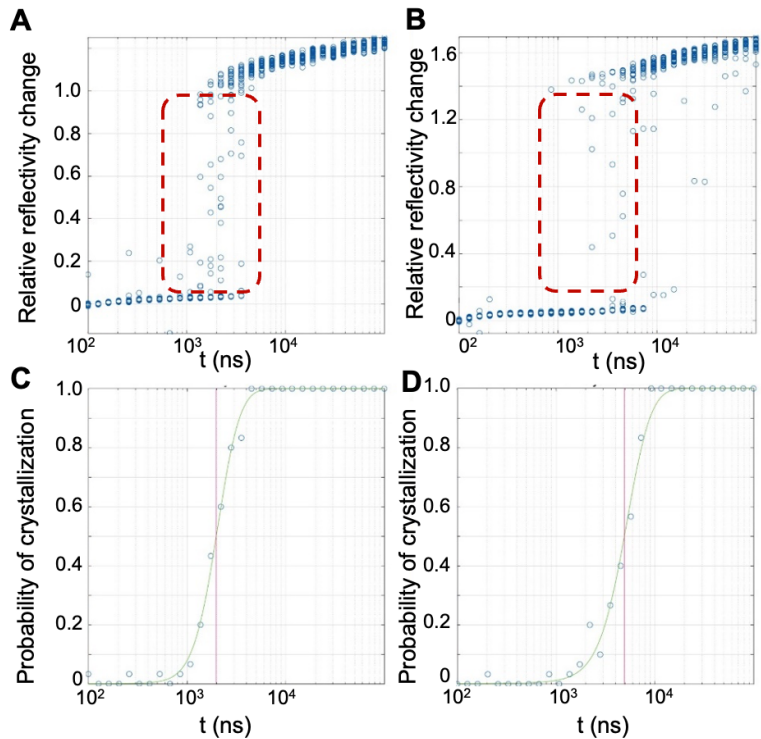

**Figure 3**: Optically induced crystallization in GeTe thin films, manifesting itself as the relative reflectivity change in the as-deposited (**A**) and the annealed (**B**) (166 °C, 3 hrs) samples as a function of laser pulse length $t$. The dashed square indicates the distribution of crystallization events within a given range of $t$. The success probability of nucleation in the as-deposited sample (**C**) and the annealed (**D**) (166 °C, 3 hrs) samples, with characteristic nucleation time $\tau$ (50% success) marked by vertical lines.

Figure 3A and B show the relative reflectivity change $\Delta R$ of as deposited and annealed (at 166°C for 3 hrs) GeTe thin films. Strong $\Delta R$ could be observed before and after the laser pulses, the higher values indicate the crystalline state of the films, while the lower ones correspond to the amorphous state close to the initial state. The regions between amorphous and crystalline represent the crystallization of the films. The minimum crystallization time $t_{min}$ and the probability of crystallization of each pulse duration can be extracted by fitting the data with the Gompertz function. As shown in Figure 3C-D, the green curves correspond to the fitted curves, and the vertical lines represent the deduced $t_{min}$ of different films. The $t_{min}$ of the as deposited film is ~1.98×10³ ns, while the value is increased to ~4.92×10³ ns for 3 hrs annealed film. Note that the $t_{min}$ measured here includes the incubation time of crystal formation, leading to larger values of minimum crystallization time than that in the devices[30–32]. Besides, the time span between 5% and 95% probability of crystallization was extended by the thermal annealing, showing ~3.06×10³ ns and 8.59×10³ ns, respectively. This suggests a more gradual change in crystallization probability with pulse duration.

In addition, within a given *t*-interval (500 s to 5000 s) (dashed square in Fig.3A-B), we observe a large number of discrete events of reflectivity changes spread over a broad range in the as-deposited sample, suggesting a stochastic nature of crystallization governed by nucleation. By contrast, the annealed sample shows less discrete events, suggesting a growth-dominated crystallization process with reduced stochasticity.

### d. Structural characterization upon annealing using synchrotron X-ray diffraction

The thermal analysis and optical measurements have clearly shown that annealing substantially slows down the crystallization kinetics. In the following, we show the results of the *in-situ* high-energy synchrotron X-ray diffraction (XRD) experiments to characterize structural evolutions in GeTe upon annealing processes.

First, we performed in-situ XRD at the DanMAX beamline of MAXIV (Sweden) with X-ray photon energy of 35 keV and monitored the structural evolution of an as-deposited sample upon continuous heating (3 K/min) up to $T_{ann}$ = 166°C (isothermal annealing temperature). As shown in Figure 4A, the sharp Bragg peaks at $Q \approx 2.1$ Å$^{-1}$ and 3 Å$^{-1}$ emerge at ~ 150°C, indicating that crystallization sets in even before $T_{ann}$ was reached. Figure 4D shows the integrated Bragg peak intensity at $Q \approx 2.1$ Å$^{-1}$, superimposed on the DSC scan of as-deposited GeTe. The rise of Bragg peak intensities approximately coincides with the stepwise drop of the exotherm at ~ 160°C in the DSC curves. This indicates that the latter is an initial step of crystallization processes. Note that no Bragg peak is observed from room temperature to ~ 157°C, indicating the exotherm at lower temperatures is related to enthalpy relaxation rather than crystallization.

Next, in-situ XRD patterns were collected during isothermal annealing for 3 hours at two selected temperatures, 132 °C and 166 °C, respectively. At 132 °C, no Bragg peaks were observed at the start of annealing, confirming that the sample remained fully amorphous prior to isothermal hold (Fig. S6); however, Bragg peaks began to emerge after 40 min of annealing. By contrast, the sample held at 166 °C exhibited Bragg peaks already at $t = 0$ (Fig. 4B), indicating that crystallization had partially occurred during the heating ramp. In both cases, the samples were partially crystallized after the isothermal annealing, which can be confirmed by the diffraction patterns collected after cooling back to room temperature (Fig. 4C).

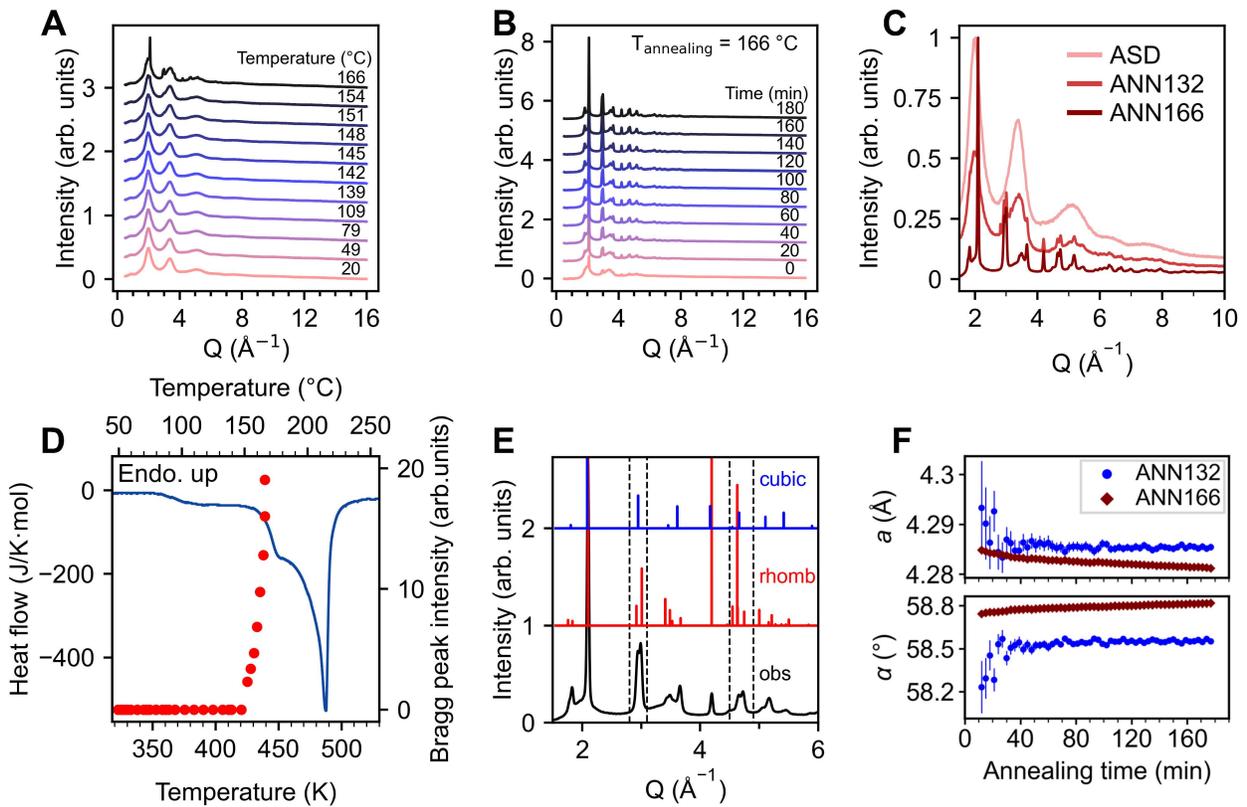

**Figure 4**: In-situ XRD characterization of GeTe during annealing at 166 °C and 132 °C. (**A**) Structural changes during heating up to 166 °C (the annealing temperature). Bragg peaks emerge at around 150 °C during heating. (**B**) In-situ XRD patterns during isothermal annealing of the as-deposited sample at 166 °C. (**C**) XRD patterns at room temperature for the samples of as-deposited, annealed at 132 °C and 166 °C, respectively. (**D**) The DSC scan (6 K/min) of the as-deposited sample, overlaid with the integrated intensity of the Bragg peak at $Q \approx 2.1$ Å$^{-1}$, indicating the onset of crystallization around 150°C. (**E**) The identification of crystallized phase after annealing at 166 °C. (**F**) The lattice parameter $a$ and α during isothermal annealing at 132 °C and 166 °C, extracted from the Rietveld refinement of the crystalline phases.

To identify the resulting crystalline phase, experimental patterns were compared with the calculated diffraction profiles of rhombohedral and cubic GeTe from the crystallographic database of ICSD (collection codes: 43202 and 202124). As shown in see Fig.4E, in both cases, the peak positions of the measured patterns are in good agreement with those of the rhombohedral phase. Particularly, the peak splitting at $Q \approx 3$ Å$^{-1}$ and 4.7 Å$^{-1}$ is consistent with the rhombohedral phase (Fig. 4E), which distinguish them from the cubic phase exhibiting single peaks. This confirms that crystallization proceeds into the rhombohedral phase, consistent with the expectation, as the cubic GeTe should only form above 700 K[33] or under pressure[34].

Having established the overall crystalline structure, we now investigate lattice parameters using Rietveld refinements. The crystalline diffraction patterns are isolated from the broad amorphous phase and the quartz capillary, where the latter two contributions are treated as the background profile. The profile of the quartz capillary is directly measured from an empty container. We use the measured data of the as-deposited sample as the profile of the amorphous phase, based on its small changes with respect to the growth of the Bragg peaks throughout the annealing process, which does not affect the refinement of the crystalline phase. The lattice parameters $a$ and α of rhombohedral GeTe, together with the scaling factors of the crystalline phase and the background contributions are optimized to achieve the best fit. The refined $a$ and α are shown in Fig. 4F. At 166 °C, the $a$ decreases by 0.09% during 180 min of isothermal annealing, accompanied by 0.10% increase in α. This gradual evolution indicates a slow relaxation of the unit cell distortion toward the cubic phase with α = 60°. By contrast, annealing at 132 °C produces an initial drop in $a$ and increase in α within the first 40 min, after which both parameters stabilize. The reasons will be discussed in Sec. Discussion.

Besides crystal structures, the structural evolution in the amorphous phase upon annealing is an interesting question to be addressed, which can be in principle analyzed through their broad scattering peaks. However, the strong intensity of the crystalline Bragg peaks, which overlapped with the amorphous peaks in the data discussed above, makes it difficult to accurately extract the structural parameters of the amorphous phase. To address this question, we annealed a sample at a lower temperature 105°C for 1 hr to avoid crystallization during annealing and performed a separate X-ray total scattering experiment with a photon energy of 60 keV at PETRAIII P02.1, DESY (Hamburg) for the as-deposited and annealed samples. Figure 5A-B and D-E show intensity $I(Q)$ patterns and the reduced pair distribution functions $G(r)$, respectively, for both samples upon heating from room temperature to 160°C. The $I(Q)$ patterns of the annealed sample do not show any Bragg peaks, indicating that no crystallization occurred during the annealing process. The first peak position of $I(Q)$, $Q_1 \sim 2$ Å$^{-1}$, shows a decreasing trend with increasing temperature (Fig. 5C), reflecting the thermal expansion. Earlier studies suggested that the temperature dependence of $Q_1^3$ in amorphous solid is inversely proportional to volume[35] $V$ through $\left(\frac{Q_1(T^0)}{Q_1(T)}\right)^3 = \frac{V(T)}{V(T^0)} = 1 + a_v(T - T^0)$, where $\alpha_v$ is the volumetric thermal expansion coefficient below $T_g$, and $T^0$ is a reference temperature. A fit to $Q_1$ yields $\alpha_V = 4.2 \times 10^{-5}$ K$^{-1}$ below 100°C for the as-deposited; and $\alpha_V = 3.7 \times 10^{-5}$ K$^{-1}$ below 100°C for the annealed sample. These values are on the same order of magnitude of those reported for crystals, $4.33 \times 10^{-5}$ K$^{-1}$ from rhombohedral single crystals[33], and $4.59 \times 10^{-5}$ K$^{-1}$ from powder diffraction[36]. Interestingly, $Q_1$ of the as-deposited sample exhibits a kink around 100°C, suggesting a faster thermal expansion ($\alpha_V = 1.4 \times 10^{-4}$ K$^{-1}$) above 100°C, while $Q_1$ of the annealed sample retains a linear trend below 120°C before merging with $Q_1$ of the as-deposited sample. Remarkably, $Q_1$ of the annealed sample is consistently lower than $Q_1$ of the as-deposited sample below 120°C, suggesting that the volume has increased after annealing. This is opposite to the behaviors of most glasses (e.g. oxide, polymers, metallic glasses), where volume often decreases after annealing below $T_g$[37,38].

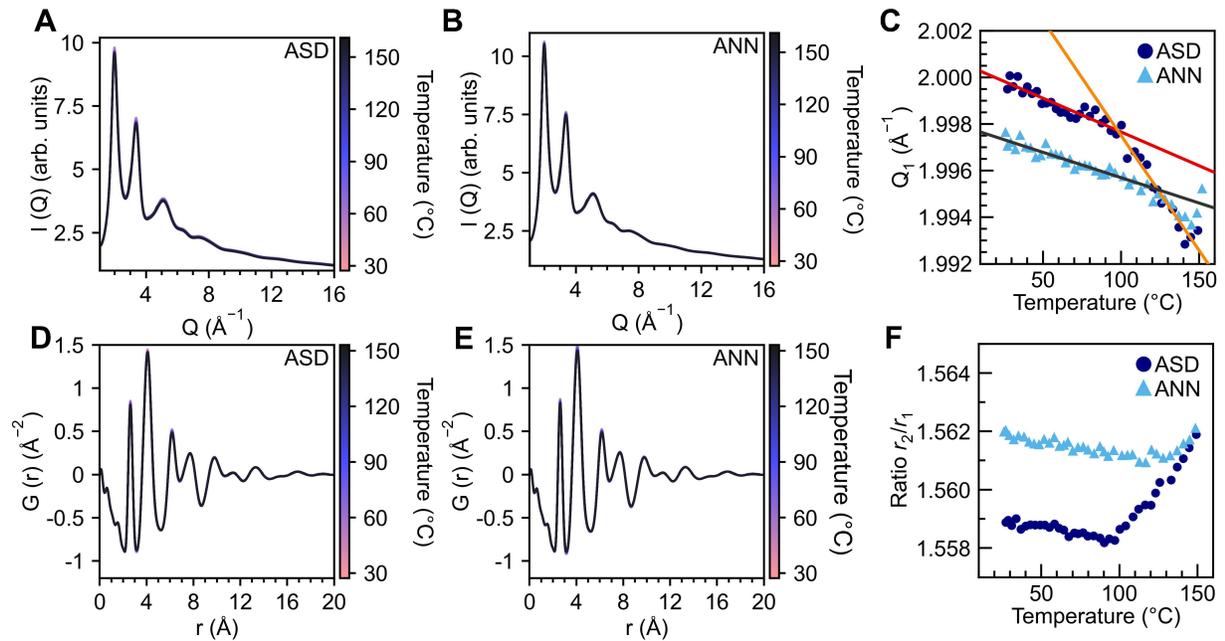

**Figure 5**: In-situ XRD during heating for as-deposited and annealed samples at 105 C for 1 hr. (**A**) Diffraction intensity patterns $I(Q)$ and (**B**) reduced pair distribution functions $G(r)$ for the as-deposited

sample. **(C)** I(Q) and **(D)** G(r) for the annealed sample. **(E)** The first peak position, $Q_1$, of I(Q) and **(F)** the ratio of the second to the first peak position of G(r), $r_2/r_1$, for both as-deposited and annealed samples.

Figure 5F shows the ratio of the second to the first peak positions of *G(r)*, $r_2/r_1$, which can serve as an indicator of Peierls-like distortions in local (defective) octahedral structural motifs in amorphous GeTe[20,34]. A larger $r_2/r_1$ value is associated with a shortening of short bonds and elongation of long bonds, reflecting larger Peierls-like distortions. The annealed sample consistently exhibits a larger $r_2/r_1$ ratio compared to the as-deposited sample before they merge above ~150°C. This suggests that annealing reinforces such distortions in amorphous GeTe, supporting the predictions from molecular dynamics simulations[20].

4. **Discussion**

The suppression of β-relaxation by thermal annealing in amorphous GeTe aligns with the behavior of a Te-free PCM $Ge_{15}Sb_{85}$[7]. Both are poor glass formers (Turnbull parameter $T_g/T_l \approx 0.47$ for GeTe and ~0.48 for $Ge_{15}Sb_{85}$) and exhibit fast crystallization rates. When the Ge:Te ratio is shifted to the eutectic composition $Ge_{15}Te_{85}$ ($T_g/T_l \approx 0.61$)[39], little annealing effect is observed due to the near absence of β-relaxation[7]. This supports the argument that in covalent glasses, β-relaxation is more tunable in PCMs than in conventional chalcogenide glasses. Both (F)DSC and optical reflectivity measurements indicate the slowdown of crystallization after annealing suppresses β-relaxations. Notably, crystallization of as-deposited GeTe exhibits a strong stochastic behavior, compared to annealed samples (Fig.3A-B). This is also evidenced by the larger error bars in $T_p$ of as-deposited samples, compared to annealed samples from FDSC measurements. This suggests that crystallization in as-deposited samples is limited by nucleation rates, whereas in annealed samples a substantial number of nuclei are already formed during the annealing process, as confirmed by the in-situ XRD data (Fig. 4). Despite the presence of numerous nuclei, crystallization progresses more slowly in the annealed samples, indicating that annealing suppresses crystal growth rates, which then become the limiting factor of the overall crystallization kinetics.

It is worth noting the lattice parameters, *a* and α, behave differently between the two partially crystallized samples (Fig. 4F). For the sample annealed at 132 °C, within the first 40 min, the rapid increase in α is likely due to the fact that the crystallization sets in during the isothermal process, where numerous small crystallites, likely nanocrystals, form. Earlier reports studied the film thickness and nanoconfinement effects on GeTe lattice parameters[40], where smaller thickness of nano crystallites leads to stronger lattice distortion (i.e. larger *c* and smaller *a* in the hexagonal representation). This is consistent with our observation of a larger lattice distortion in the initial stage of crystallization (see Fig.S6). As these crystallites grow, their lattice parameters converge toward those of bulk crystals. By comparison, the sample annealed at 166 °C has already crystallized during the heating ramp, such that its lattice parameters measured at 166 °C correspond to bulk crystals.

The increased ratio $r_2/r_1$ after annealing is consistent with the trend observed in amorphous $Ge_{15}Sb_{85}$. In both compositions, *ab initio* molecular dynamics simulations showed two types of locally favorable structures: (defective) octahedral and tetrahedral motifs. In principle, both Peierls-like distortions in octahedral motifs and a larger fraction of tetrahedral motifs may lead to larger values of $r_2/r_1$. In $Ge_{15}Sb_{85}$, the Peierls-like distortions are reinforced, and the tetrahedral fraction reduces upon annealing[7]. In GeTe, with a higher Ge-atomic concentration, Raty et al.[20] showed that Ge-atoms form homopolar bonds, resulting in tetrahedral motifs. Although the tetrahedral motifs fraction is likely higher in GeTe, compared to $Ge_{15}Sb_{85}$, annealing below $T_g$ also reduces tetrahedral motifs[20]. Therefore, the increased $r_2/r_1$ should be attributed to reinforced Peierls-like distortions. Peierls-like distortions are associated with higher covalency and more localized electrons[12]. The reduction of Peierls-like distortions by high pressures is related to the closing of pseudo-gap in density of state and possibly a semiconductor-to-metal transition[34]. Persch et al.[12], increasing Peierls-like distortions through varying pseudo-binary compositions from GeTe to GeSe, showed that a higher covalency composition $GeTe_{0.3}Se_{0.7}$ exhibits about four orders of magnitude slower crystallization kinetics, compared to GeTe. In this context, the reinforced Peierls-like distortions in GeTe corresponds to more covalent bonding with rigid network, accommodating fewer local fast dynamics in the glass and thus suppressed β-relaxations. Earlier studies of metallic, molecular and polymer glasses also demonstrated that annealing may suppress the β-relaxation, although it often takes substantially longer time[41–43]. We note a remarkable difference: in most other glasses, volume decreases upon annealing below $T_g$, which are intuitively related to dense atomic packing and the vanishing of the local fast mobility regions. However, in GeTe, the lower $Q_1$ after annealing, suggesting an increased volume, implies that volume or density is not the crucial factor for reducing local dynamics in GeTe.

## 5. Conclusion

The suppressed β-relaxation implies a reduction of local fast mobility regions in glass dynamics, associated with reinforced Peierls-like distortions in amorphous structure. The effects on crystallization in GeTe resembles that in $Ge_{15}Sb_{85}$ and can be understood through classical nucleation and growth theory, where both the nucleation and growth rates are limited by kinetic factors (atomic mobility). A suppressed β-relaxation results in lower atomic mobility. The slowdown effect on crystallization is present even after annealing introduces nuclei and partial crystallization. Recent molecular dynamics simulations of simple glasses showed that the activation of β-relaxations corresponds to a percolation transition[44], where atoms clusters percolate, enabling long range mass transport. It is possible that a similar percolation transition underlies the β-relaxations in PCMs, such as GeTe. Machine learned potentials recently developed for PCMs would be promising to simulate glass dynamics, where long time dynamics and large box models are required in those covalently bonded systems and clarify their atomic-level mechanism.

**Acknowledgement**

We acknowledge the MAX IV Laboratory for beamtime on the DanMAX beamline under proposal 20220516. Research conducted at MAX IV, a Swedish national user facility, is supported by Vetenskapsrådet (Swedish Research Council, VR) under contract 2018-07152, Vinnova (Swedish Governmental Agency for Innovation Systems) under contract 2018-04969 and Formas under contract 2019-02496. DanMAX is funded by the NUFI grant no. 4059-00009B. The authors thank N. Amini for assistance during the DanMAX beamtime. We acknowledge DESY (Hamburg, Germany), a member of the Helmholtz Association HGF, for the provision of experimental facilities. Parts of data were collected using PETRAIII P02.1. The authors thank YD Cheng for preliminary PETRAIII data analysis. The project is supported by the Deutsche Forschungsgemeinschaft (DFG, German Research Foundation) (AOBJ: 670132). We thank the Villum Fonden (42116), Novo Nordisk Fonden (NNF21OC0071257) and the Danish Agency for Science, Technology, and Innovation for funding the instrument center DanScatt.

(AOBJ: 670132). We thank the Villum Fonden (42116), Novo Nordisk Fonden (NNF21OC0071257) and the Danish Agency for Science, Technology, and Innovation for funding the instrument center DanScatt.

**Competing interests**

The authors declare no competing interests.

**Differential scanning calorimetry measurements**

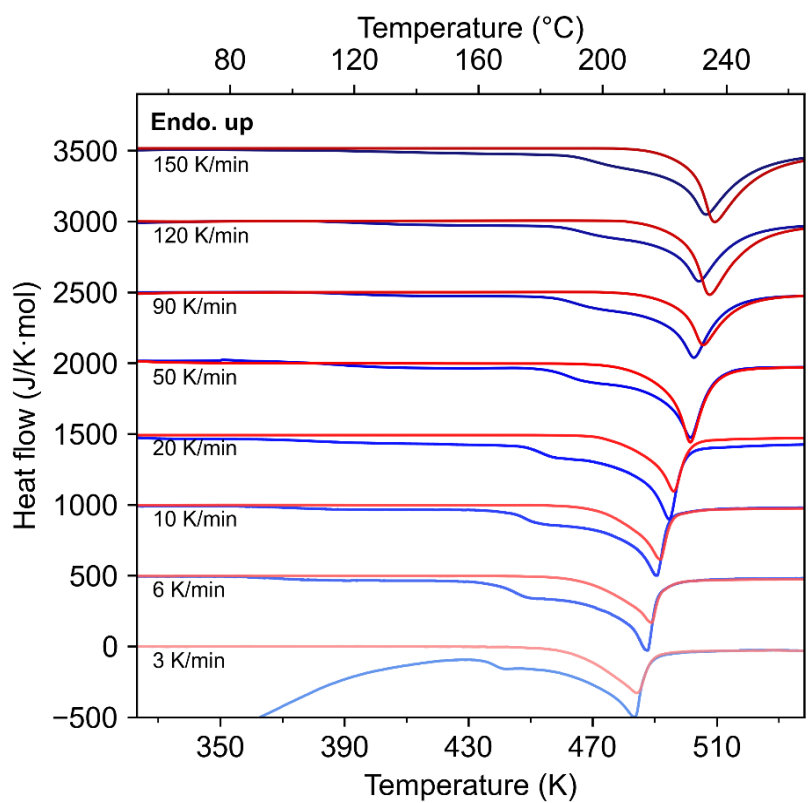

Figure S1. Differential scanning calorimetry (DSC) thermograms of GeTe: as-deposited (blue) and annealed (red) measured at heating rates of 3–150 K·min$^{-1}$ (labeled). Temperature is shown in K (bottom) and °C (top); endothermic up. Curves are vertically offset pairwise for clarity.

**Kissinger plot numerical simulation**

From the crystalline volume fraction model for isothermal crystallization, it is possible to simulate crystallization peak temperatures under constant heating rates, as previously demonstrated for Ge2Sb2Te5 [1]. In the present work, an analogous approach was employed to simulate the Kissinger plot. The governing equations and the key aspects of the simulation procedure are summarized below.

Rate constant k(T) by Avrami equation

$$k(T) = k_\infty \cdot exp(-E_K / (k_B \cdot T)) \qquad (1)$$

where

$k_\infty$: prefactor (from Kissinger fitting of experimental data),

$E_K$: activation energy (from Kissinger),

$k_B$: Boltzmann constant,

$T$: temperature.

Crystalline volume fraction (by JMAK for isothermal crystallization)

$$X(t) = 1 - exp(-(k(T) \cdot t)^n) \qquad (2)$$

with X(t) ∈ [0,1] the crystalline volume fraction, $t$ = time, and $n$ the Avrami exponent.

Artificial time

Define the artificial time (evaluated at temperature T) as

$$t_a(X)|_T = (1 / k(T)) \cdot ( ln(1 / (1 - X)) )^{(1/n)} \qquad (3)$$

This is the time needed to reach the current X value at the new temperature T.

Simulation setup

Let

$\Delta T$: temperature increment,

$q$: heating rate,

$\Delta t = \Delta T / q$: time increment.

We discretize the temperature range into steps of $\Delta T$ and iterate over index $i = 0,1,2,\ldots$

Initial step ($i=0$)

$T(0) = T_0$, $k(0) = k(T_0)$, $t_a(0) = 0$, $X(0) = 0$

Update for *i > 0*

$T(i) = T(i-1) + \Delta T, \quad k(i) = k(T(i))$

$$t_a(i) = (1 / k(T(i))) \cdot ( \ln(1 / (1 - X(i-1))) )^{(1/n)} \quad (4)$$

$$X(i) = 1 - \exp( -( k(T(i)) \cdot (\Delta t + t_a(i)) )^n ) \quad (5)$$

The parameters used in the simulation are summarized in Table 1.

| Sample and heating rate range | $E_K$ [J] | $k_\infty$ [1/s] fitted | $k_\infty$ [1/s] adjusted | $n$ |
|---|---|---|---|---|
| ASD low  | $6.90 \cdot 10^{-19}$ | $5.74 \cdot 10^{42}$ | $7 \cdot 10^{42}$ | 5.8 |
| ASD high | $2.38 \cdot 10^{-19}$ | $1.34 \cdot 10^{16}$ | $5 \cdot 10^{16}$ | |
| ANN low  | $5.35 \cdot 10^{-19}$ | $4.70 \cdot 10^{32}$ | $9 \cdot 10^{32}$ | |
| ANN high | $2.16 \cdot 10^{-19}$ | $1.95 \cdot 10^{14}$ | $7.5 \cdot 10^{14}$ | |

Table 1. Summary of crystallization parameters employed in the Kissinger plot simulation. ASD refers to as-deposited samples and ANN to annealed samples. The low heating-rate range corresponds to 20–1000 K/s, while the high heating-rate range covers 1000–16 000 K/s. The activation energy $E_K$ and pre-exponential factor $k_\infty$ were obtained from linear fitting of experimental Kissinger plots. Since the intercept was not described accurately by the fit, k∞ values were manually adjusted. The Avrami exponent *n* was adopted from [2].

Crystalline volume fraction was simulated over the temperature range 300–998.15 K (melting point). The calculation was terminated once the volume fraction reached 0.999 999 (nearly full crystallization). The crystallization peak temperature was then identified as the maximum of the derivative of the crystalline volume fraction with respect to temperature.

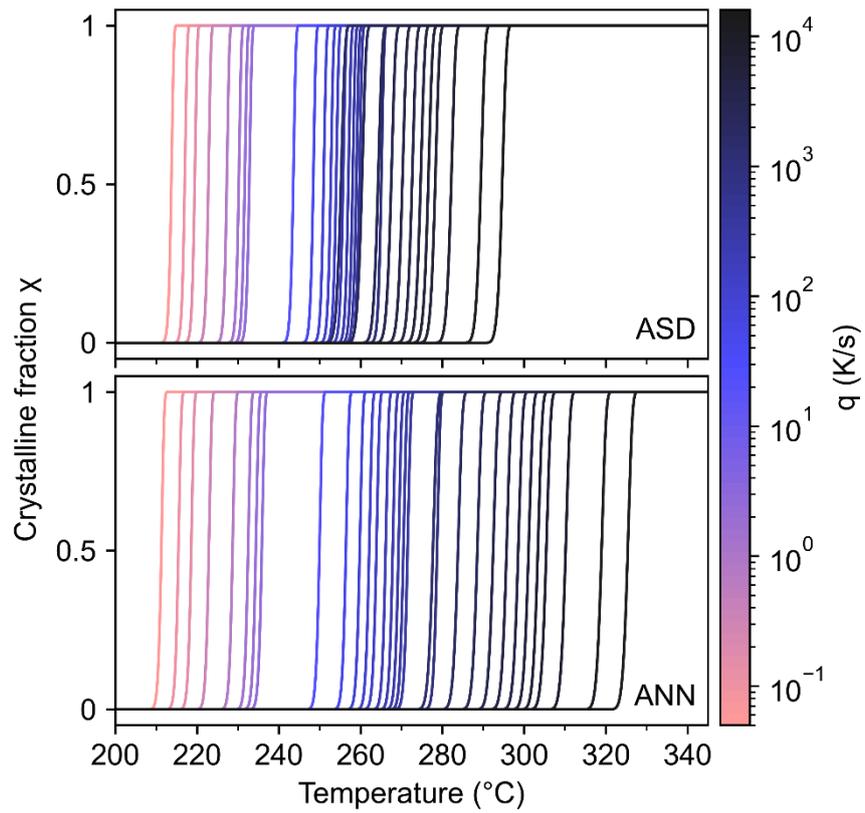

Figure S2. Simulated crystalline volume fraction X of GeTe for (top) as-deposited (ASD) and (bottom) annealed (ANN) samples at different heating rates q. The crystalline fraction is shown as a function of temperature, with color indicating the applied heating rate from 0.05 to 16 000 K/s.

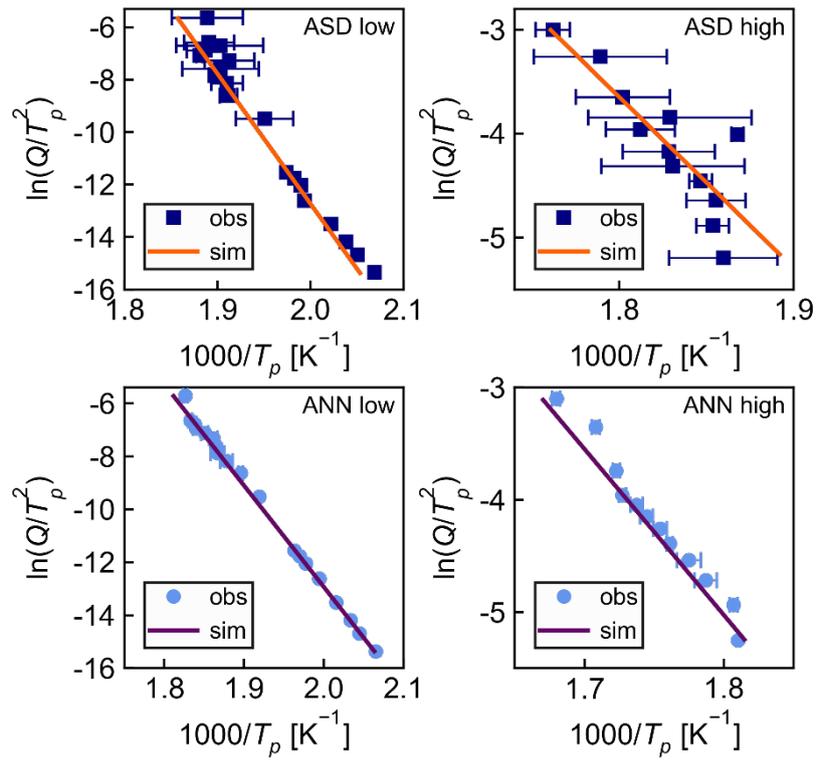

Figure S3. Kissinger plots of crystallization peak in GeTe for as-deposited (ASD, top) and annealed (ANN, bottom) samples at low (left) and high (right) heating rates. Symbols show observed data (obs) with error bars, and solid lines represent simulated trends (sim).

*In-situ* X-ray diffraction

Crystallization onset

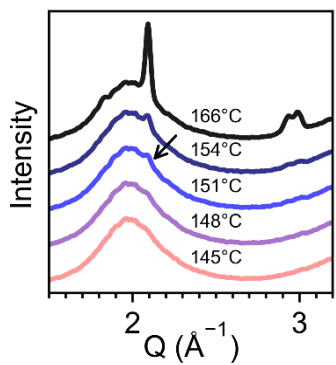

Figure S4. High temperature XRD patterns of amorphous GeTe during heating. The most prominent Bragg peak begins to emerge at 151 °C, marking the onset of crystallization.

Crystalline phase identification

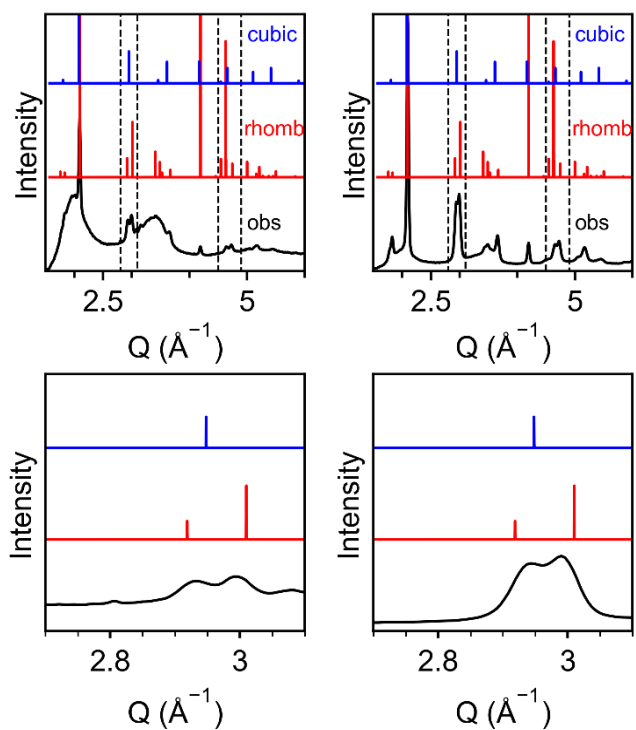

Figure S5. X-ray diffraction patterns of GeTe after isothermal annealing at **132 °C (left)** and **166 °C (right)**. The full Q-range (top panels) is shown together with simulated cubic (blue) and rhombohedral (red) reference patterns. The magnified regions (bottom panels) highlight the characteristic peak splitting of the rhombohedral phase, enabling identification of the crystalline structure.

<u>Isothermal annealing at *T=132°C* and lattice parameter changes in hexagonal notation</u>

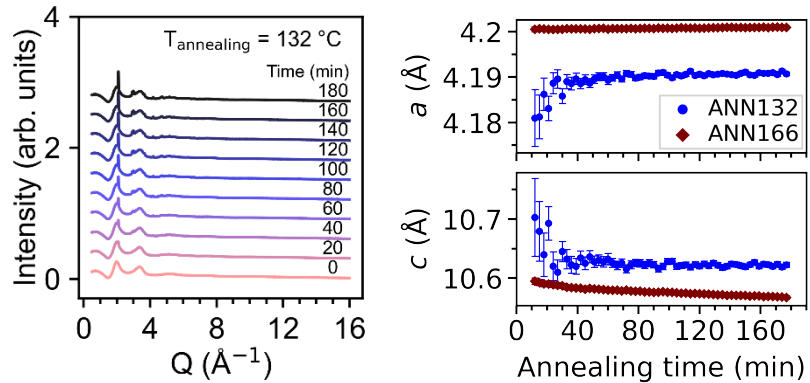

Figure S6. GeTe during isothermal annealing. Left: *in-situ* XRD patterns at 132 °C; vertical offset is applied for clarity, and numbers indicate elapsed annealing time (min). Right: crystalline lattice parameters a and c (hexagonal notation) versus annealing time for samples annealed at 132 °C (blue circles) and 166 °C (red diamonds).